# Generalized vibration curves for apodization and super-resolution in far-field diffraction

Lowell T. Wood

Department of Physics, University of Houston, Houston, Texas 77204-5005

## Abstract

The Cornu spiral, widely used to analyze Fresnel diffraction, is a familiar example of a vibration curve, also known as an amplitude-phase curve or phasor diagram. A related construction appears in discussions of Fraunhofer diffraction, where the aperture function is uniform, making the curve's shape depend solely on phase. When the aperture function varies with position, however, both amplitude and phase determine the geometry.

Here, vibration curves are extended to continuous, monotonic, symmetric, nonuniform aperture functions that produce either apodization (broadening of the central maximum with suppression of side maxima) or super-resolution (narrowing of the central maximum with enhancement of side maxima). These generalized vibration curves provide an intuitive visualization of how aperture structure shapes the far-field diffraction pattern. They clarify the mechanisms behind apodization and super-resolution and reveal how phasor rotation at the first diffraction zero provides a signature that characterizes the two effects. This unified perspective serves as a useful pedagogical tool for connecting Fourier transforms, optical diffraction, and vibration curves in the undergraduate and graduate curriculum.

## Introduction

The study of apodization[1–17] and super-resolution[9–16] (sometimes called inverse apodization) has a long history in optics, yet standard treatments often give only brief attention to their physical and mathematical basis. Apodization broadens the central diffraction maximum and suppresses secondary maxima, while super-resolution sharpens the central peak at the cost of stronger side lobes. Both demonstrate how changes in an aperture function can dramatically reshape a diffraction pattern.

Apodization and super-resolution arise in a wide range of wave-based systems beyond classical diffraction experiments. In Fourier transform spectroscopy and signal processing, window functions are routinely applied to suppress ringing and spectral leakage associated with finite data records[18]. In optical imaging and astronomical instrumentation, pupil apodization is used to control diffraction sidelobes and improve contrast, particularly where diffraction sidelobes might obscure faint companions in binary stars and exoplanet imaging[19,20]. Closely related ideas appear in antenna arrays, ultrasound imaging, and magnetic resonance imaging, where amplitude and phase weighting across an array or aperture are used to tailor beam shapes, suppress sidelobes, or produce super-resolving responses[21]. In all these cases, the underlying physics is governed by the coherent superposition of complex field contributions, with amplitude weighting and phase structure determining how interference redistributes energy in the resulting pattern.

Super-resolution techniques, in particular, rely critically on engineered phase variations to delay destructive interference and sharpen selected features of the diffraction pattern, necessarily at the expense of increased sidelobes and reduced robustness[22].

Experimental demonstrations and introductory simulations using nonuniform aperture functions were reported previously[14,15], but that work did not emphasize the deeper connection between the aperture function and the resulting diffraction structure. The present work focuses instead on a unified geometric interpretation using generalized vibration curves.

The literature generally treats this connection indirectly: Jacquinot and Roizen-Dossier[12] offered a detailed discussion of apodization but without vibration curve analysis. Goodman[9] briefly noted its physical origin but did not pursue the connection further. Lipson *et al*.[23] noted that curvature depends on the aperture function, though they did not develop the idea. Finally, Bracewell[24] indicated that this type of curve is a useful tool for qualitative thinking and some numerical work but did not treat the idea formally.

Although classic treatments stopped short of vibration-curve analysis, the Cornu spiral, also known as the Euler spiral and appearing in contexts ranging from elasticity to railroad transition curves as well as Fresnel diffraction, serves as a special case of such a curve.[25] The present work extends this idea to generalized vibration curves constructed directly from the aperture function, showing how amplitude and phase contributions combine to shape the far-field diffraction pattern. This work also shows that the geometry of these generalized vibration curves contains simple angular signatures of apodization and super-resolution. Closely tied to Fourier analysis and the mathematics of diffraction, vibration curves provide an intuitive and visually engaging framework for understanding how aperture weighting redistributes energy in the diffraction pattern. They are especially useful as a teaching aid, offering students a compact and accessible way to connect mathematical analysis with physical intuition.

Although the examples presented here are restricted to real, symmetric amplitude apertures for pedagogical clarity, the vibration-curve approach readily extends to apertures with spatially varying phase or complex transmission functions, without requiring additional theory.

In what follows, we first review the concept of vibration curves and show how they can be generalized for arbitrary aperture functions. We then provide motivation for choosing these examples, describe how the vibration curves were created, give information concerning their interpretation, present details that illustrate apodization and super-resolution, and show how the aperture function affects the diffraction amplitude for each case.

## Theory

In its simplest form, the one-dimensional diffraction integral may be written as

$$E(x) = \int_{-\infty}^{\infty} A(x') e^{i\phi(x',x)}\, dx', \tag{1}$$

where $A(\mathrm{x}')$ is the aperture function (real) that sets the limits of the integration, $\phi(\mathrm{x}', \mathrm{x})$ is the phase produced by the geometry and/or by a phase mask, and $E(x)$ is the scalar electric field produced at the diffraction plane. Primed variables denote the aperture plane; unprimed variables denote the observation plane. Here, the obliquity (inclination) factor and any terms outside the integral that affect only the amplitude of the diffraction pattern are neglected. Figure 1 shows the geometry for the one-dimensional diffraction calculation. This development follows very closely that of Towne[26], except here aperture functions that are not uniform are considered. This addition creates some interesting changes in the types of vibration curves that are produced.

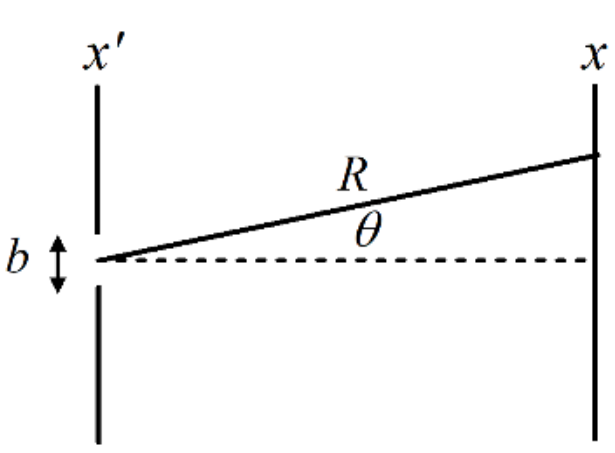

Fig. 1 One-dimensional diffraction geometry. The aperture width is *b*. *R* and $\theta$ determine the observation point.

Define

$$Z(u) = \int_{o}^{u} A(u') e^{i\phi(u')} du'. \tag{2}$$

If $Z(u) = X(u) + iY(u)$, then

$$X(u) = \int_{0}^{u} A(u') \cos[\phi(u')] du' \tag{3}$$

and

$$Y(u) = \int_{0}^{u} A(u') \sin[\phi(u')] du'. \tag{4}$$

Here $X(u)$ and $Y(u)$ are the *x*- and *y*-components in the complex plane and represent the real and imaginary parts of the electric field, respectively.

The relationship between $u'$ and $x'$ is established once $\phi(x, x')$ is given. The expressions for the differential arc length $ds$, slope of the curve $\frac{dY}{dX}$, and curvature $K$ are

$$ds = [\mathrm{d}X^2 + \mathrm{d}Y^2]^{\left(\frac{1}{2}\right)} = A(u)\mathrm{d}u, \tag{5}$$

$$\frac{\mathrm{d}Y}{\mathrm{d}X} = \tan[\phi(u)], \tag{6}$$

independent of the form of $\phi(u)$, and

$$K = \frac{\mathrm{d}\phi}{\mathrm{d}s} = \frac{[A(u)]^{-1}\,\mathrm{d}\phi}{\mathrm{d}u}. \tag{7}$$

This curvature dependence on the inverse of the aperture function is the key element in understanding the mathematical reasons for apodization and super-resolution. It is worth mentioning that the product of the curvature and arc length $\mathrm{d}s$ is a constant $\mathrm{d}\boldsymbol{\phi}$. This must be the case because only far-field diffraction is considered, and the linear phase results in a constant phase difference between the phasors.

Consider

$$Z_{21} = Z_2 - Z_1 = \int_{u_1}^{u_2} A(u')e^{i\phi(u')}du' = \int_{0}^{u_2} A(u')e^{i\phi(u')}du' - \int_{0}^{u_1} A(u')e^{i\phi(u')}du'. \quad (8)$$

A graph of $Y(u)$ versus $X(u)$ traces a vibration curve with $|Z_{21}|^2$ giving the square of the chord length between $(X_1, Y_1)$ and $(X_2, Y_2)$ which, in turn, is proportional to the intensity of the diffraction pattern. $u$ is the arc length along the curve, rescaled by the aperture function according to Eq. (5). For cases in this work, however, only diffraction amplitudes are considered, so the length need not be squared.

**Creating and interpreting the vibration curves**

The aperture functions examined in this paper were selected to illustrate distinct and physically meaningful ways in which amplitude weighting across an aperture influences the diffraction amplitude and its geometric representation using vibration curves. The selection was guided both by previous experimental results and by the desire to isolate specific physical effects rather than to provide an exhaustive survey of possible apodization and super-resolution schemes.

Two of the aperture functions, the cosine (apodizing) and partial inverted cosine (super-resolving), were chosen to connect directly with experimental results reported in earlier work [14]. Including these functions allows a comparison with other aperture functions that illustrate important differences in vibration curve structure.

The triangular aperture function (apodizing) was included as a familiar reference case whose Fourier transform is nonnegative for all spatial frequencies, making it a useful contrast to other aperture functions that show no such behavior. Its vibration curve exhibits additional looping associated with the nonnegative nature of its Fourier transform.

The full inverted cosine aperture was selected to contrast directly with the partial inverted cosine. By keeping the functional form fixed while increasing the strength of the amplitude variation across the aperture, this comparison demonstrates the role of weighting strength in producing super-resolution effects.

The inverted triangle function was chosen to compare its behavior to that of the full inverted cosine aperture, and the resulting behavior is qualitatively similar.

Figures 2 and 3 show the graphs of the normalized aperture functions (left) and their corresponding normalized diffraction amplitudes (right). In the aperture function graphs, the solid vertical lines mark the edge of the aperture, and in the diffraction curves, the solid vertical lines mark the first zeros of the uniform (Fraunhofer) diffraction amplitude for comparison with the zeros of other aperture functions.

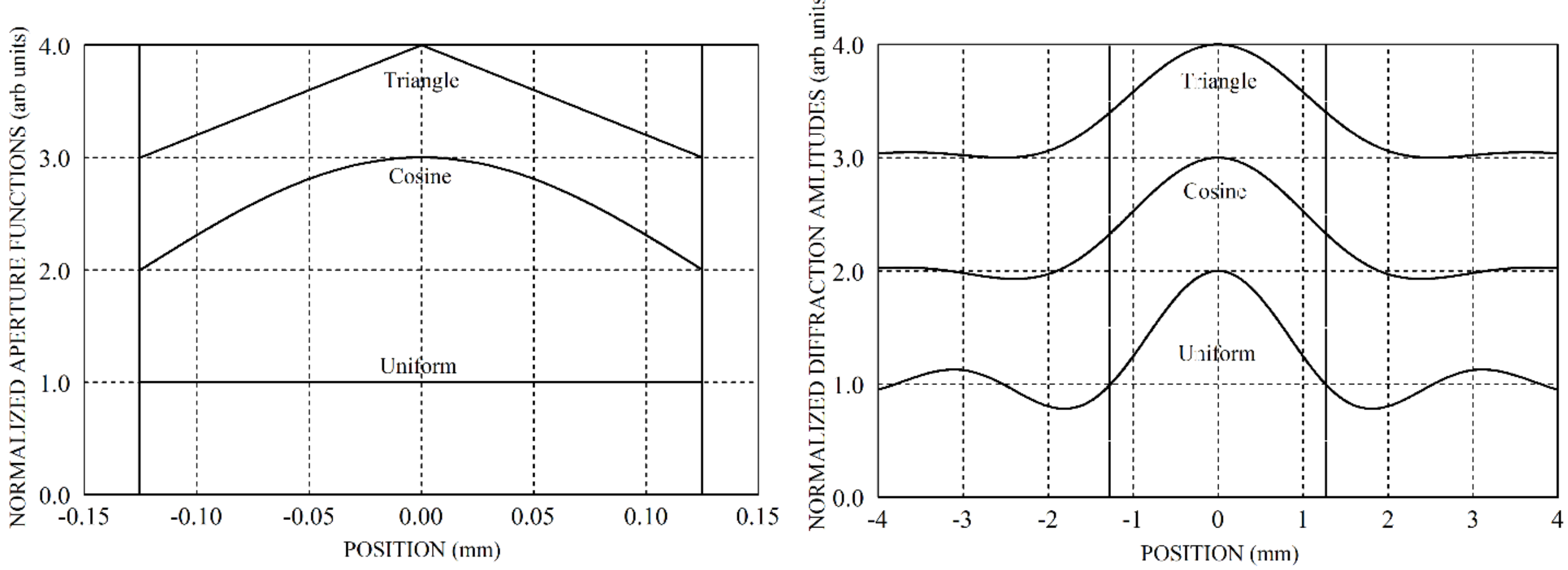

Fig. 2. Normalized apodizing aperture functions (left) and their corresponding normalized diffraction amplitude functions (right) for the cosine, triangle, and uniform apertures. The uniform aperture is shown for comparison purposes, and the two solid vertical lines mark the zeros of the diffraction amplitude. Vertical zeros are displaced by one arbitrary unit. The aperture-to-screen distance $R$ is 500 mm for all cases in this work.

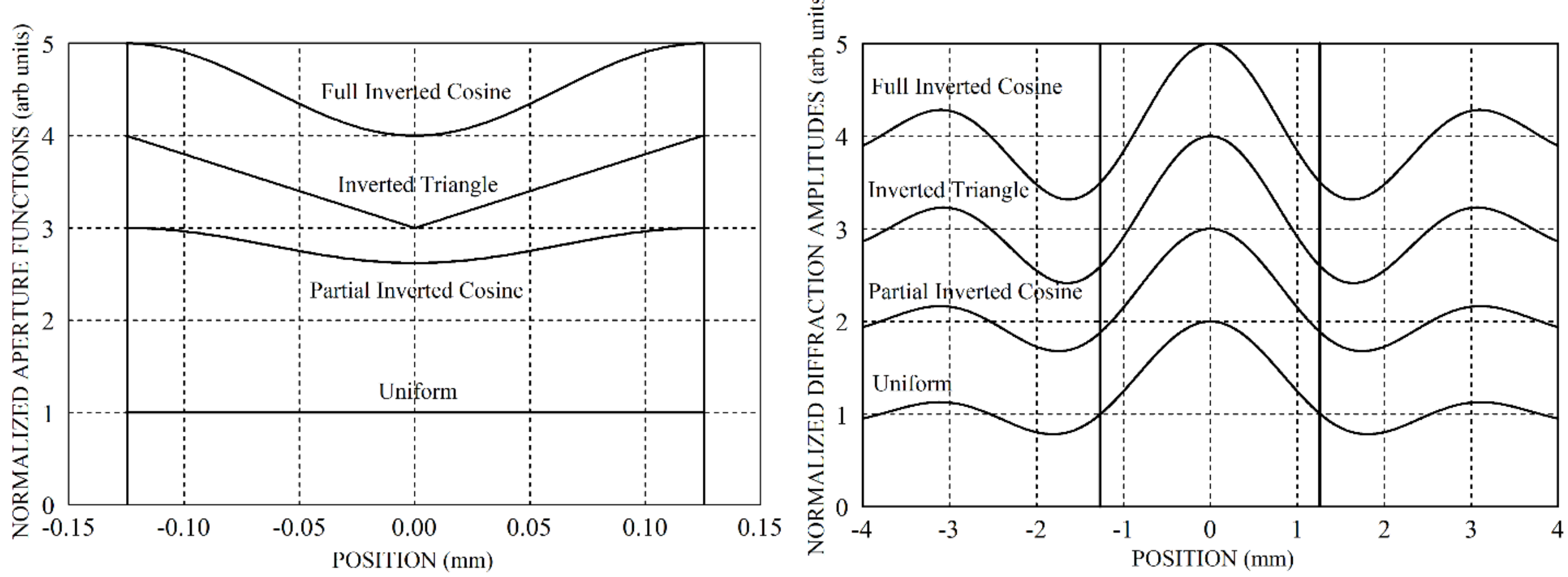

Fig. 3. Normalized super-resolving aperture functions (left) and their corresponding normalized diffraction amplitude functions (right) for the partial inverted cosine, inverted triangle, full inverted cosine, and uniform apertures. The uniform aperture is shown for comparison purposes, and the two solid vertical lines mark the zeros of the diffraction amplitude. Vertical zeros are displaced by one arbitrary unit.

Table I provides a summary of the functional form, type and name, qualitative behavior, and diffraction amplitude expression.

| Aperture Function A(x') | Type and Name | Diffraction Behavior | Diffraction Amplitude Expression |
|---|---|---|---|
| A(x') = 1 | Uniform | Standard sinc pattern | $b\ \mathrm{sinc}((kb\sin\theta)/2)$ |
| A(x') = cos(πx'/b) | Apodizing Cosine function | Wider central maximum, lower side lobes | $\frac{(2\pi b\cos((kb\sin\theta)/2))}{(\pi^2 - (kb\sin\theta)^2)}$ |
| A(x') = 1 - \|2x'/b\| | Apodizing Triangle function | Smooth profile, sinc² behavior | $b/2\ \mathrm{sinc}^2((kb\sin\theta)/4)$ |
| A(x') = A₀ - ε cos(πx'/b) | Super-Resolving Partial inverted cosine | Moderate sharpening and increased side lobes | $\frac{2A_o\sin(kb\sin\theta/2)}{k\sin\theta} - \frac{2\pi b\epsilon\cos(kb\sin\theta/2)}{\pi^2 - k^2b^2\sin^2\theta}$ |
| A(x') = \|2x'/b\| | Super-Resolving Inverted triangle | Strong narrowing, large side lobes | $\frac{2\sin(kb\sin\theta/2)}{k\sin\theta} - \frac{8\sin^2(kb\sin\theta/4)}{k^2b\sin^2\theta}$ |
| A(x') = 1 - cos(πx'/b) | Super-Resolving Full inverted cosine | Strong narrowing, large side lobes | $\frac{2\sin(kb\sin\theta/2)}{k\sin\theta} - \frac{2\pi b\cos(kb\sin\theta/2)}{\pi^2 - k^2b^2\sin^2\theta}$ |

Table I. Aperture function designation, name of function (apodizing or super-resolving), diffraction behavior, and diffraction amplitude expression. Here $A_o = 0.81$ and $\epsilon = 0.19$.

The diffraction patterns and corresponding vibration curves were calculated using an aperture width $b$ of 0.25 mm and an aperture-to-observation distance $R$ of 500 mm. A helium-neon laser (wavelength 632.8 nm) was the illuminating source. These parameters place the system in the Fraunhofer (far-field) regime, where the phase difference between each phasor is the same for each aperture function for a given observation angle $\theta$. The phase factor, $\phi(\mathrm{x}',\mathrm{x}) = kx'\sin\theta,$ is written in terms of the observation angle θ using a sine function rather than spatial frequency, since this notation is more familiar to most students. Integration of the real and imaginary parts of $A(x')e^{ikx'\sin\theta}$ may be done over each segment of the aperture, but rather than evaluating a separate integral for each aperture segment, a cumulative integral was first evaluated across the aperture using

$$X(x') = \int_0^{x'} A(x'')\cos(kx''\sin\theta)\,dx'' \text{ and } Y(x') = \int_0^{x'} A(x'')\sin(kx''\sin\theta)\,dx''.$$

By taking advantage of the symmetry of the aperture functions, the vibration curve is constructed by sampling this cumulative integral at adjacent aperture positions, starting at the lower edge of the aperture at $-0.125$ mm and proceeding to the upper edge at 0.125 mm in steps of 0.005 mm.

Each segment between adjacent points corresponds to a differential contribution, and the cumulative vector from point $i$ to point $f$ is the net electric field. The vibration curve is referenced so that the initial point corresponds to the lower edge of the aperture, and the final point corresponds to the upper edge. Once the $X$ and $Y$ values for the vibration curve are known, the curves were then plotted for characteristic values of $\theta$ corresponding to the diffraction maximum $A_o$ (shown in black), fractions of the maximum (2/3 and 1/3) (shown in red and green, respectively), the first zero (shown in purple), and the first secondary maximum or minimum (shown in blue). These color codes apply to all continuous vibration curves in this paper.

Continuous vibration curves are labeled with $i$ (initial point) and $f$ (final point). The curves are to be traversed in the counter-clockwise direction with each segment of the curve representing the amplitude and phase of the electric field contribution at that observation angle. The solid black vector represents the diffraction amplitude at the angle $\theta$ for which the vibration curve was calculated. Closed curves (shown in purple) correspond to zero amplitude.

As a reminder of the more conventional case, Fig. 4 shows the vibration curves for Fraunhofer diffraction. The vibration curves are always arcs of circles or complete circles, depending on the angle at which the vibration curve is evaluated. For the zero curve (purple), the phasor starts at the top and turns counterclockwise through 360 degrees when it closes to give zero.

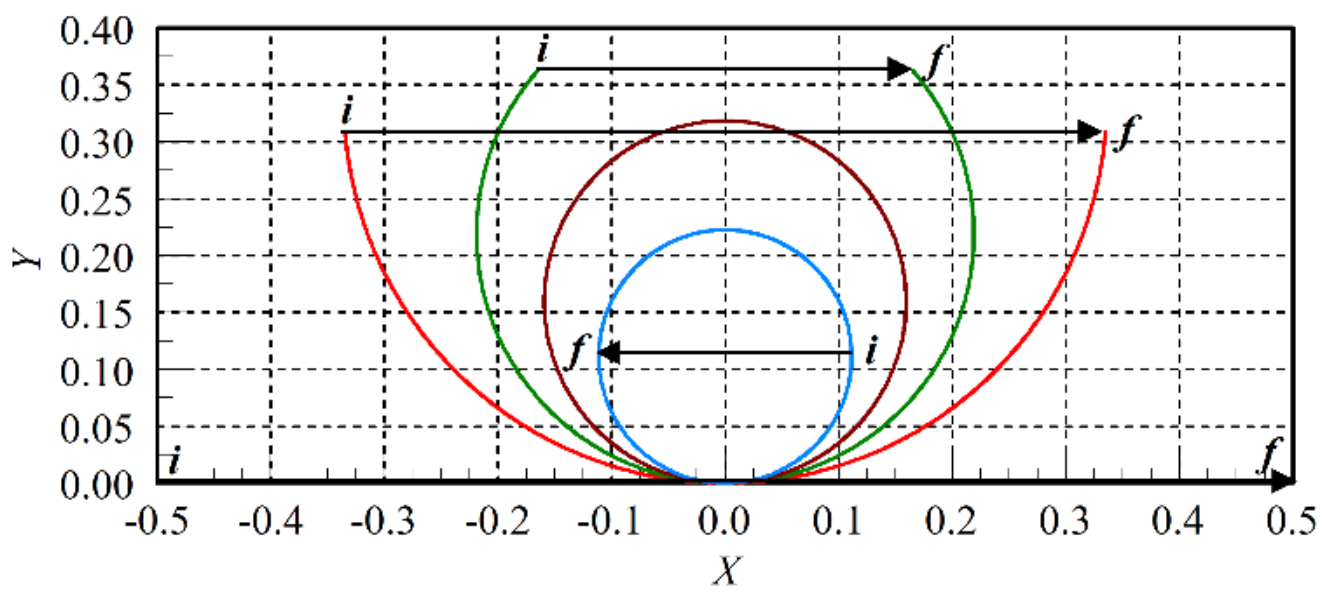

Fig. 4. (Color online) Normalized Fraunhofer vibration curves for $A_o$ (black), (2/3) $A_o$ (red), (1/3) $A_o$ (green), first zero (purple), and first secondary maximum (blue). The blue curve is one and one-half circles. All are circles or arcs of circles.

**Results and discussion: vibration curve examples**

We now examine representative vibration curves to illustrate how apodizing and super-resolving aperture functions differ geometrically and physically. Figures 5 and 6 present the continuous vibration curves for the apodizing cases. Unlike the Fraunhofer case, these curves are not circular arcs but display continuously varying curvature that reflects the position-dependent weighting of the aperture function. The cosine aperture function produces a cardioid-like curve at the first zero, with the phasor turning through an angle greater than 360° before folding back to reach the first minimum. The triangular aperture function overlaps itself even at the first zero and again at the subsequent maximum, consistent with the positivity of its Fourier transform.

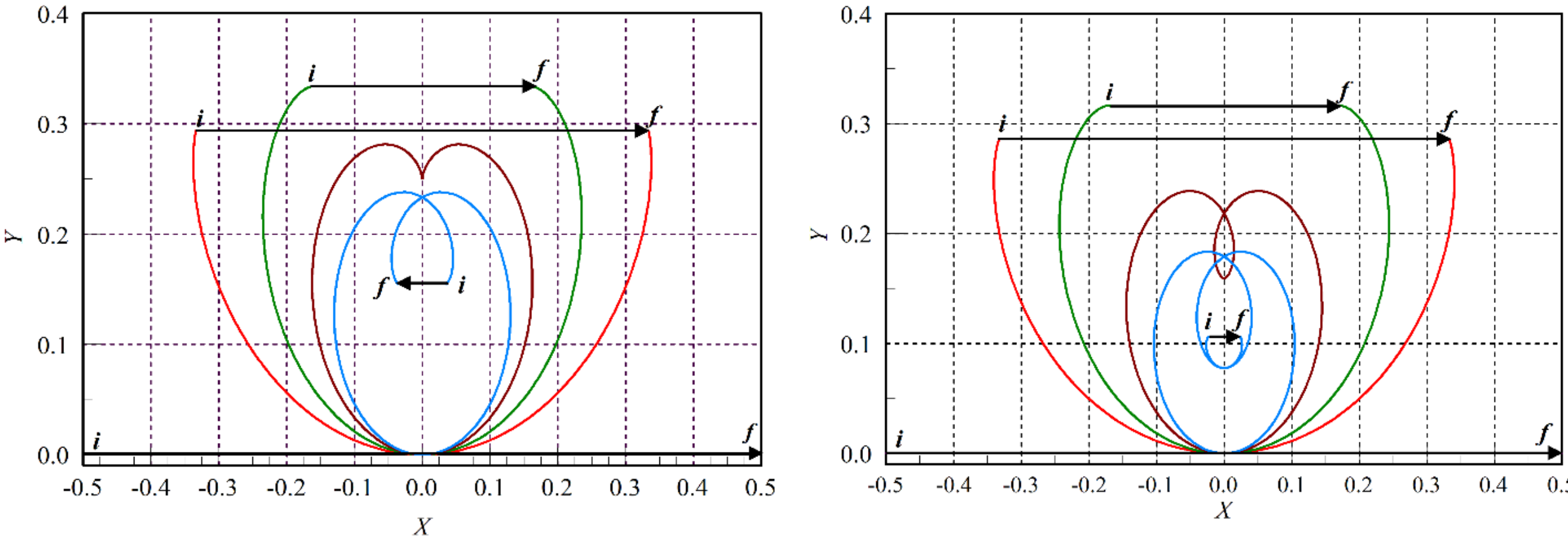

Fig. 5. (Color online) Normalized cosine vibration curves for $A_o$ (black), (2/3) $A_o$ (red), (1/3) $A_o$ (green), first zero (purple), and first minimum (blue).

Fig. 6. (Color online) Normalized triangle vibration curves for $A_o$ (black), (2/3) $A_o$ (red), (1/3) $A_o$ (green), first zero (purple), and first secondary maximum (blue). Note that the direction of the arrow is positive.

Figures 7 and 8 show the discrete phasor diagrams corresponding to the first zero and the first minimum of the cosine aperture function. These diagrams clearly illustrate how the position dependence of the aperture function influences the local curvature, and they make more transparent the angle through which the phasor must turn to close or reach a minimum.

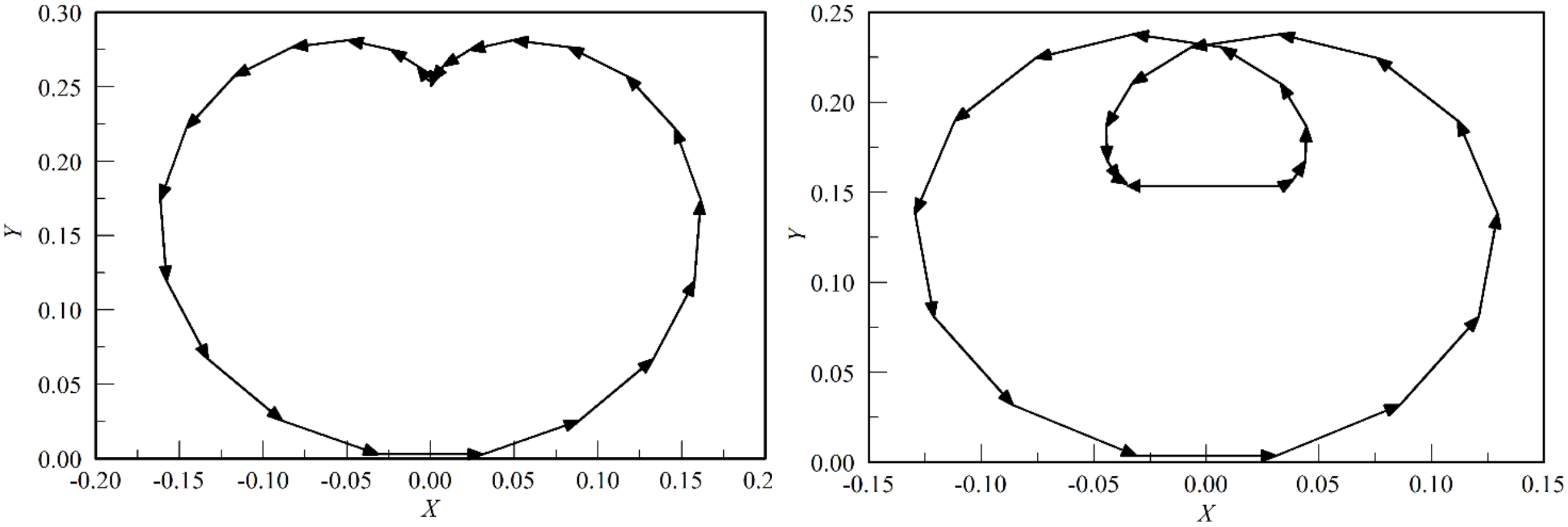

Fig. 7. Discrete cosine vibration curve for the first diffraction zero. Note the decreasing radius of curvature as the edge contribution closes the curve at the top.

Fig. 8. Discrete cosine vibration curve for the first diffraction minimum. As the minimum is reached, the radius of curvature decreases.

In the super-resolving cases (Figs. 9–11), the curvature becomes extreme near the aperture center because of weak amplitude weighting at that point. The vibration curves bend sharply and overshoot, accounting for the intensified secondary extrema characteristic of super-resolution. Among these, the partial inverted cosine produces the mildest super-resolving effect due to its less pronounced curvature profile.

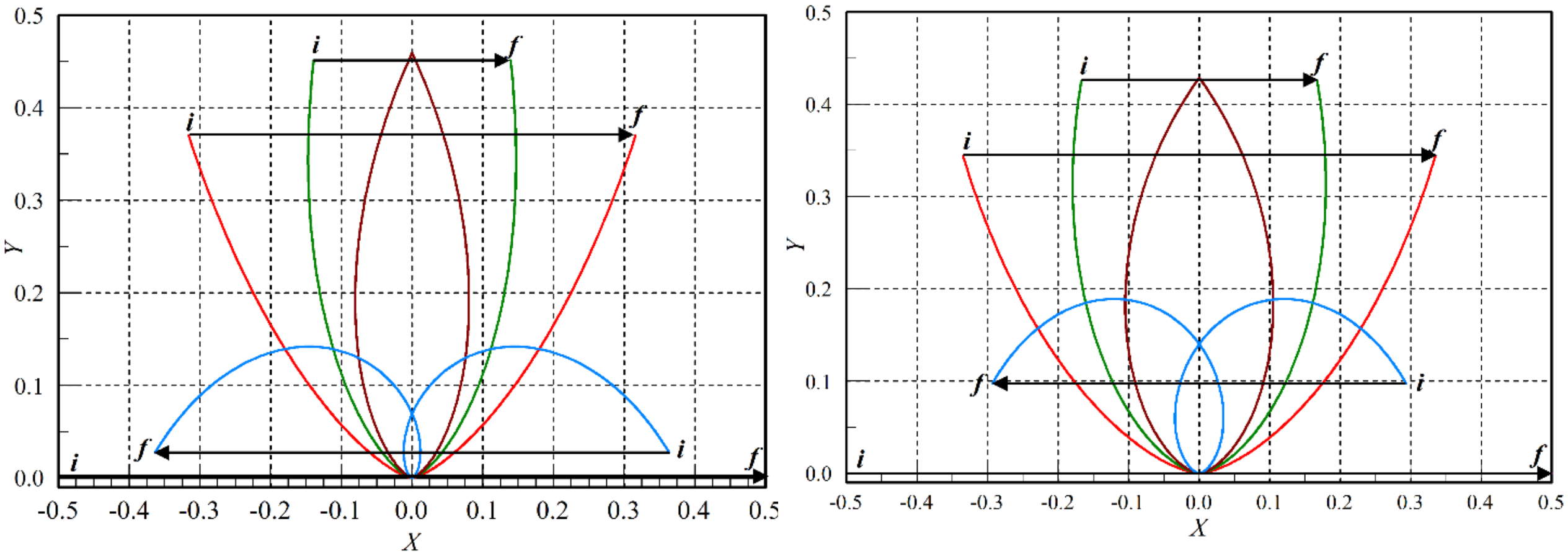

Fig. 9. (Color online) Normalized full inverted cosine vibration curves $A_o$ (black), (2/3) $A_o$ (red), (1/3) $A_o$ (green), first zero (purple), and first minimum (blue). Note the very large secondary minimum.

Fig. 10. (Color online) Normalized inverted triangle vibration curves $A_o$ (black), (2/3) $A_o$ (red), (1/3) $A_o$ (green), first zero (purple), and first minimum (blue). Note the very large secondary minimum.

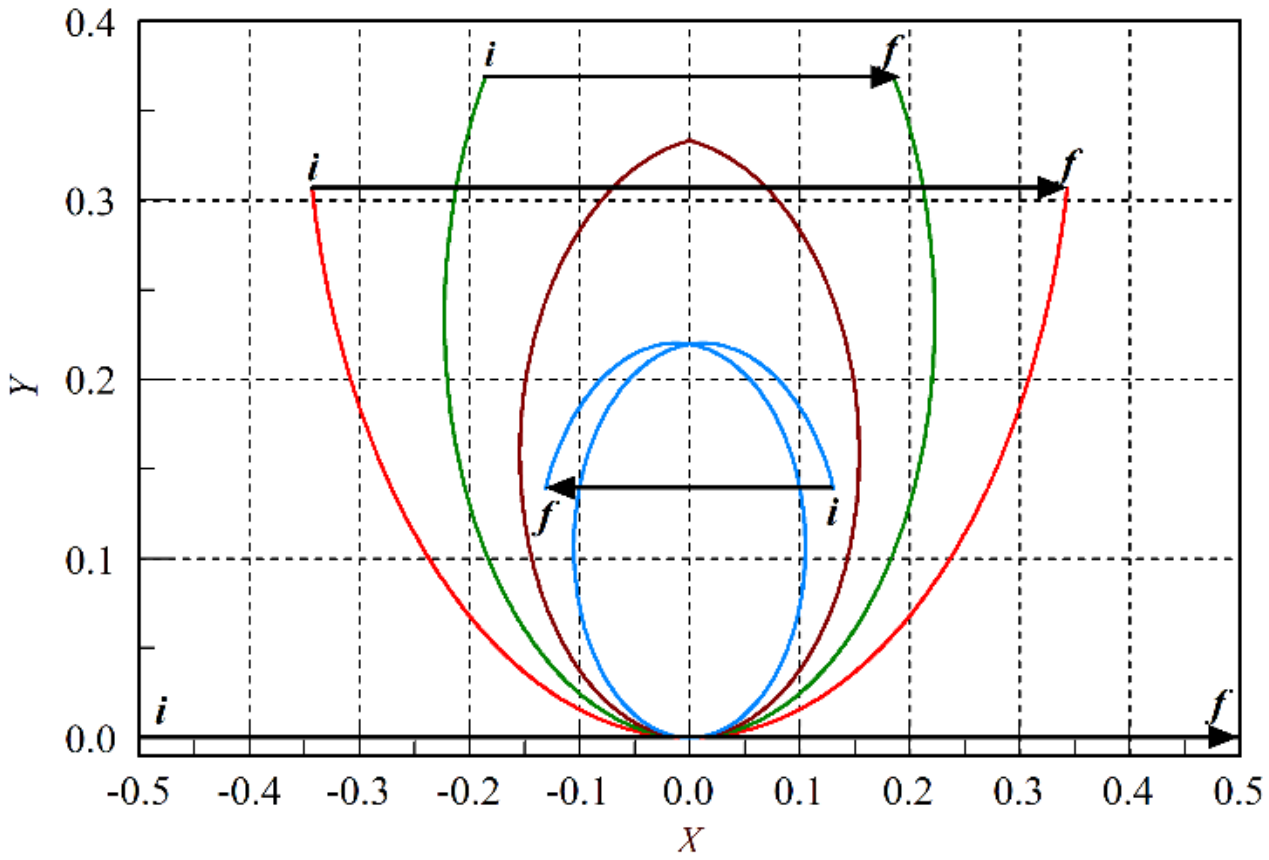

Fig. 11. (Color online) Normalized partial inverted cosine vibration curves $A_o$ (black), (2/3) $A_o$ (red), (1/3) $A_o$ (green), first zero (purple), and first minimum (blue). Note the reduced degree of super-resolution.

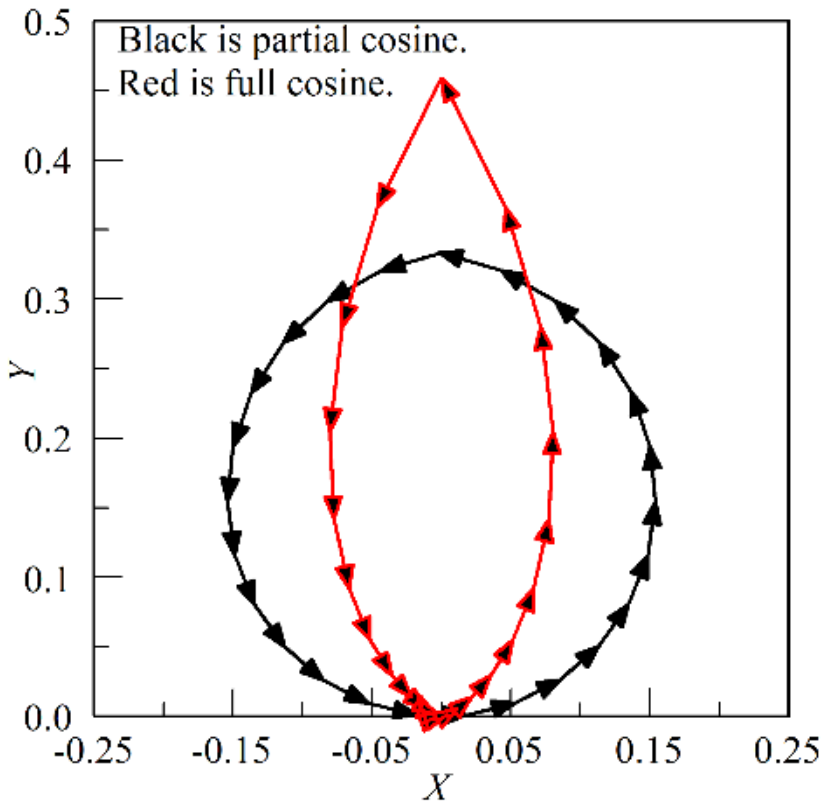

Fig. 12. (Color online) Normalized partial inverted cosine and full inverted cosine discrete vibration curves. The full inverted cosine curve (red) has a much smaller radius of curvature at the center that causes the phasor to close more quickly.

Discrete phasor diagrams further clarify these effects by showing how variations in phasor length control the curvature of the vibration path. When the change from center to edge of the aperture is small, the degree of apodization or super-resolution is correspondingly reduced. This trend is evident in Fig. 12, which compares the vibration curves at the first zero for the full inverted cosine and the partial inverted cosine apertures. Super-resolution is clearly diminished when the aperture variation is less extreme.

Finally, Fig. 13 shows the normalized inverted triangle vibration curve at the first minimum. As the phasors from the right side (negative part of the aperture) add, their lengths shorten and the radius of curvature decreases, becoming very small when the central phasors contribute. The resulting sharp overshoot, driven by this small radius of curvature, produces a large first minimum in the diffraction amplitude and a narrower central diffraction maximum.

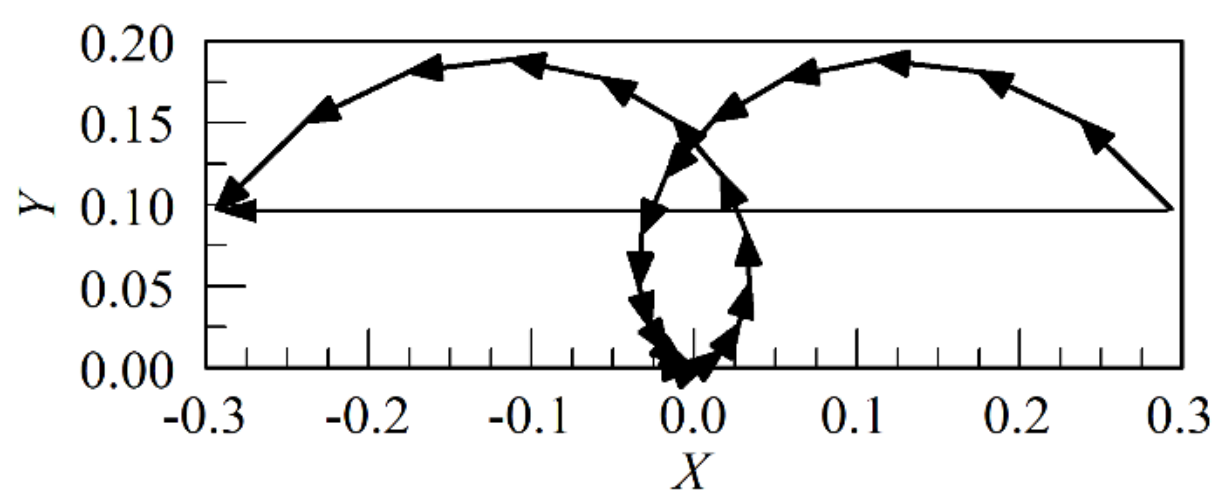

Fig. 13. Inverted triangle function vibration curve at the first minimum. The very small radius of curvature caused by phasors at the center of the aperture produces the large overshoot and subsequent large first minimum.

These examples demonstrate that the geometry of vibration curves captures in a direct and visual way the essential differences between uniform, apodizing, and super-resolving apertures. These geometric signatures, particularly the total phasor rotation, connect aperture weighting to diffraction behavior.

**Conclusions**

By analyzing five nonuniform, continuous, symmetric aperture functions, this work demonstrates that generalized vibration curves are a powerful way to visualize how apodization and super-resolution arise from the aperture's structure. These curves clearly show how every part of the aperture function contributes to every part of the diffraction pattern when the point density is sufficiently high, and they may be viewed as elementary building blocks from which more complex apodizing or super-resolving profiles can be constructed. Particularly important is the curvature change introduced by the nonuniform aperture function amplitude.

The geometry of the curve contains a simple and general signature: the net angular sweep of the phasor from the initial point $i$ to the final point $f$ at the first diffraction zero. For the uniform aperture, this sweep is exactly 360°, increasing beyond 360° for apodization and decreasing below 360° for super-resolution. This angular property, derived entirely from the aperture function, provides a direct visual and quantitative link between aperture weighting and the redistribution of energy in the diffraction pattern.

Apodization arises when the aperture function decreases monotonically from the center to the edge, whereas super-resolution arises when it increases. The degree to which each effect occurs depends on the magnitude of this change, which directly governs how strongly energy is shifted between the central maximum and the side lobes.

Because the diffraction amplitude is the Fourier transform of the aperture function to within a phase factor, vibration curves offer an intuitive bridge between the abstract mathematics of Fourier analysis and the physical intuition central to optics. They complement the Cornu spiral tradition and provide a compact, visually engaging teaching aid for illustrating how changes in an aperture function can dramatically reshape the diffraction pattern. Vibration curves show that the shape of the aperture function is written directly into the geometry of diffraction, making them an especially effective tool for both teaching and conceptual understanding.

The author has no conflicts of interest.